\documentstyle[12pt]{article}
\begin{document}

%
%
\def\PRL#1#2#3{{\sl Phys. Rev. Lett.} {\bf#1} (#2) #3}
\def\NPB#1#2#3{{\sl Nucl. Phys.} {\bf B#1} (#2) #3}
\def\NPBFS#1#2#3#4{{\sl Nucl. Phys.} {\bf B#2} [FS#1] (#3) #4}
\def\CMP#1#2#3{{\sl Commun. Math. Phys.} {\bf #1} (#2) #3}
\def\PRD#1#2#3{{\sl Phys. Rev.} {\bf D#1} (#2) #3}
\def\PRv#1#2#3{{\sl Phys. Rev.} {\bf #1} (#2) #3}
\def\PLA#1#2#3{{\sl Phys. Lett.} {\bf #1A} (#2) #3}
\def\PLB#1#2#3{{\sl Phys. Lett.} {\bf #1B} (#2) #3}
\def\JMP#1#2#3{{\sl J. Math. Phys.} {\bf #1} (#2) #3}
\def\JNMP#1#2#3{{\sl J. Nonl. Math. Phys.} {\bf #1} (#2) #3}
\def\PTP#1#2#3{{\sl Prog. Theor. Phys.} {\bf #1} (#2) #3}
\def\SPTP#1#2#3{{\sl Suppl. Prog. Theor. Phys.} {\bf #1} (#2) #3}
\def\AoP#1#2#3{{\sl Ann. of Phys.} {\bf #1} (#2) #3}
\def\PNAS#1#2#3{{\sl Proc. Natl. Acad. Sci. USA} {\bf #1} (#2) #3}
\def\RMP#1#2#3{{\sl Rev. Mod. Phys.} {\bf #1} (#2) #3}
\def\PR#1#2#3{{\sl Phys. Reports} {\bf #1} (#2) #3}
\def\AoM#1#2#3{{\sl Ann. of Math.} {\bf #1} (#2) #3}
\def\UMN#1#2#3{{\sl Usp. Mat. Nauk} {\bf #1} (#2) #3}
\def\FAP#1#2#3{{\sl Funkt. Anal. Prilozheniya} {\bf #1} (#2) #3}
\def\FAaIA#1#2#3{{\sl Functional Analysis and Its Application} {\bf #1} (#2)
#3}
\def\BAMS#1#2#3{{\sl Bull. Am. Math. Soc.} {\bf #1} (#2) #3}
\def\TAMS#1#2#3{{\sl Trans. Am. Math. Soc.} {\bf #1} (#2) #3}
\def\InvM#1#2#3{{\sl Invent. Math.} {\bf #1} (#2) #3}
\def\LMP#1#2#3{{\sl Letters in Math. Phys.} {\bf #1} (#2) #3}
\def\IJMPA#1#2#3{{\sl Int. J. Mod. Phys.} {\bf A#1} (#2) #3}
\def\IJMPB#1#2#3{{\sl Int. J. Mod. Phys.} {\bf B#1} (#2) #3}
\def\AdM#1#2#3{{\sl Advances in Math.} {\bf #1} (#2) #3}
\def\RMaP#1#2#3{{\sl Reports on Math. Phys.} {\bf #1} (#2) #3}
\def\IJM#1#2#3{{\sl Ill. J. Math.} {\bf #1} (#2) #3}
\def\APP#1#2#3{{\sl Acta Phys. Polon.} {\bf #1} (#2) #3}
\def\TMP#1#2#3{{\sl Theor. Mat. Phys.} {\bf #1} (#2) #3}
\def\JPA#1#2#3{{\sl J. Physics} {\bf A#1} (#2) #3}
\def\JSM#1#2#3{{\sl J. Soviet Math.} {\bf #1} (#2) #3}
\def\MPLA#1#2#3{{\sl Mod. Phys. Lett.} {\bf A#1} (#2) #3}
\def\JETP#1#2#3{{\sl Sov. Phys. JETP} {\bf #1} (#2) #3}
\def\CAG#1#2#3{{\sl  Commun. Anal\&Geometry} {\bf #1} (#2) #3}
\def\JETPL#1#2#3{{\sl  Sov. Phys. JETP Lett.} {\bf #1} (#2) #3}
\def\PHSA#1#2#3{{\sl Physica} {\bf A#1} (#2) #3}
\def\PHSD#1#2#3{{\sl Physica} {\bf D#1} (#2) #3}
\def\PJA#1#2#3{{\sl Proc. Japan. Acad.} {\bf #1A} (#2) #3}
\def\JPSJ#1#2#3{{\sl J. Phys. Soc. Japan} {\bf #1} (#2) #3}
\def\SJPN#1#2#3{{\sl Sov. J. Part. Nucl.} {\bf #1} (#2) #3}
\def\JKPS#1#2#3{{\sl J. Korean Phys. Soc. } {\bf #1} (#2) #3}
\def\JPIV#1#2#3{{\sl J. Phys. IV} {\bf #1} (#2) #3}
\def\ibid#1#2#3{{\sl ibid} {\bf #1} (#2) #3}

\begin{titlepage}
\vspace{.2in}
\begin{center}
{\large\bf Affine Toda model coupled to matter and the string tension in QCD$_{2}$}
\end{center}

\vspace{1in}


\begin{center}

Harold Blas

\vspace{.5 cm}
\small

\par \vskip .1in \noindent
Instituto de F\'\i sica Te\'orica - IFT/UNESP\\
Rua Pamplona 145\\
01405-900  S\~ao Paulo-SP, BRAZIL\\
E-mail: blas@ift.unesp.br

\normalsize
\end{center}


\vspace{1 cm}

\begin{abstract}
The $sl(2)$ affine Toda model coupled to matter (ATM) is shown to describe various features, such as the spectrum and string tension, of the low-energy effective Lagrangian of QCD$_{2}$ (one flavor and $N$ colors). The corresponding string tension is computed when the dynamical quarks are in the {\sl fundamental} representation of $SU(N)$ and in the {\sl adjoint} representation of $SU(2)$.  
\end{abstract}





\par \vskip .3in \noindent
\vspace{1 cm}
\end{titlepage}

It has been conjectured that the low-energy action of QCD$_{2}$ ($e>>m_{q}$, $m_{q}$ quark mass and $e$ gauge coupling) might be related to massive two dimensional integrable models, thus leading to the exact solution of the strong coupled QCD$_{2}$ \cite{frishman}. Although some hints toward an integrable structure in QCD$_{2}$ have been encountered the problem remains open \cite{abdalla}. 

In recent papers by Armoni et al. \cite{armoni} it was proved, that bosonized QCD$_{2}$ \cite{frishman} exhibits a screening nature (vanishing of the string tension) when the dynamical quarks have no mass both in the case when the source and the dynamical fermions belong to the same representation of the gauge group and in the case when the representation of the external charge is smaller than the representation of the massless fermions. The string tension also vanishes when the test charges are in the adjoint representation and the dynamical ones in the fundamental representation. Confinement is restored in the non-standard matter content case (e.g., dynamical adjont matter and fundamental probe charge) when a small mass ($m_{q}<<e$) is given to the quarks, as initially argued in \cite{gross}. Similar phenomena occur in QED$_{2}$ \cite{abdalla}. Integer charges can screen fractional charges  when the dynamical electrons are massless. The confinement phase is restored when the dynamical electrons are massive and when the external charge is not an integer multiple of the dynamical charge. The string tension in  QCD$_{2}$ is \cite{armoni}
\begin{eqnarray}
\label{tension}
\sigma\,=\, m_{q} \mu_{R} \sum_{i} \Big[ 1- \mbox{cos}\, 4\pi \lambda_{i} \frac{k_{ext}}{k_{dyn}}\Big],
\end{eqnarray}
where $\mu_{R} \sim e$\, ($\mu_{fund}=\frac{\mbox{exp}(\gamma)}{(2 \pi)^{3/2}} e$ , \,$\gamma$ is the Euler number), $\lambda_{i}$ are the isospin eigenvalues of the dynamical representation, $k_{ext}$ and $k_{dyn}$ are the affine current algebra levels of the external and dynamical representations, respectively. $R=$ fundamental and adjoint representations. A possible generalization of (\ref{tension}) to representations to which the bosonization techniques are applicable, among them the antisymmetric and symmetric representations, is outlined  in \cite{armoni}. 

Besides, the $sl(n)$ affine Toda model coupled to matter (Dirac) fields (ATM) \cite{nucl}-\cite{matter} constitute excellent laboratories to test ideas about confinement \cite{nucl1}, the role of solitons in quantum field theories \cite{nucl}, duality transformations interchanging solitons and particles \cite{nucl, jmp}, as well as the reduction processes of the (two-loop) WZNW theory from which the ATM models are derivable \cite{matter, topological, annals}. 

We show that the $sl(2)$ ATM model describes the low-energy spectrum of QCD$_{2}$ (one flavor and $N$ colors in the fundamental and $N=2$ in the adjoint representations, respectively). The exact computation of the string tension is performed. A key r\^ole will be played by the equivalence between the Noether and topological currents at the quantum level \cite{nucl1}.

The Lagrangian of the ATM model is \cite{nucl, nucl1}\cite{annals}
\begin{eqnarray}
\label{atm}
 \frac{1}{k}{\cal L} = -\frac{1}{4} \partial_{\mu} \varphi \, \partial^{\mu} \varphi 
+ i  {\bar{\psi}} \gamma^{\mu} \partial_{\mu} \psi
- m_{\psi}\,  {\bar{\psi}} \,
e^{2i\varphi\,\gamma_5}\, \psi,
\end{eqnarray}
where $k=\frac{\kappa}{2\pi}$,\, ($\kappa \in \hbox{\sf Z}$), $\varphi$ is a real field, $m_{\psi}$ is a mass parameter, and $\psi$ is a Dirac spinor. Notice that ${\bar{\psi}} \equiv {\widetilde{\psi}}^{T} \,\gamma_0$. We shall take ${\widetilde{\psi}}=e_{\psi} \psi^{*}$ \cite{nucl1}, where $e_{\psi}$ is a real dimensionless constant. 
The conformal version (CATM) of (\ref{atm}) has been constructed in \cite{matter}. The integrability properties and the reduction processes: WZNW$\rightarrow$ CATM $\rightarrow$ ATM $\rightarrow$ sine-Gordon(SG) $+$ free field, have been considered \cite{nucl, topological, annals} \cite{nucl1}. The $sl(n)$ ATM exhibits a generalized sine-Gordon/massive Thirring correspondence \cite{jmp}. Moreover, (\ref{atm}) exhibits mass generation despite chiral symmetry \cite{witten} and confinement of fermions in a self-generated potential \cite{nucl1, chang}.

The Lagrangian is invariant under $\varphi \rightarrow \varphi + n \pi$, thus the topological charge, $Q_{\mbox{topol.}} \equiv \int \, dx \, j^0 ,
\,j^{\mu} =  \frac{1}{\pi}\epsilon^{\mu\nu} \partial_{\nu} \, \varphi$, can assume nontrivial values. A reduction is performed imposing the constraint \cite{nucl1, annals} \cite{nucl}
\begin{eqnarray}
\label{equiv}
\frac{1}{2\pi}\epsilon^{\mu\nu} \partial_{\nu} \, \varphi=
\frac{1}{\pi} \bar \psi \gamma^\mu  \psi,
\end{eqnarray}
where $J_{\mu}= \bar \psi \gamma^\mu  \psi$ is the $U(1)$ Noether current.

The Eq. (\ref{equiv}) implies $\psi^{\dagger} \psi \sim \partial_{x} \varphi$,  thus the Dirac field is confined to live in regions where the field  $\varphi$ is not constant. The $1(2)-$soliton(s) solution(s) for $\varphi$ and $\psi$ are of the sine-Gordon (SG) and massive Thirring (MT) types, respectively; they satisfy (\ref{equiv}) for $|e_{\psi}| = 1$, and so are solutions of the reduced model \cite{nucl1}. Similar results hold in $sl(n)$ ATM \cite{bueno, jmp}.

Introduce a new boson field representation of fermion bilinears as \cite{stone}\begin{eqnarray}
: {\bar{\psi }}(1\pm \gamma _{5})\psi : \,=\, 
- \frac{c \mu}{\pi} :  e^{(\pm i \sqrt{4 \pi} \phi)} :  ,\,\,\,\,\,\,\,
: {\bar{\psi }}\gamma ^{\mu }\psi : \,=\,-\frac{1}{\sqrt{\pi}}
\epsilon^{\mu \nu}\partial_{\nu }\phi  
\label{boson1},
\end{eqnarray}
where $c=\frac{1}{2}\exp{(\gamma)}$ and $\mu$ is an infrared regulator. Define the fields $\Phi$ and $\rho$ as
\begin{eqnarray}
\label{linear}
\Phi = \frac{2}{\beta} (\sqrt{\pi} \phi + \varphi);\,\,\,\,\, \rho = \frac{\sqrt{2/\pi}}{\beta} (2\sqrt{\pi}e_{\psi} \phi + \pi \varphi).
\end{eqnarray}

Then the Lagrangian (\ref{atm}) becomes \cite{nucl1}
\begin{eqnarray}
{\cal L}_{bos} = - \frac{\epsilon}{2 e_{\psi}}  (\partial_{\mu} \rho )^2+ \epsilon\, \frac{1}{2} (\partial_{\mu }\Phi )^{2}
+  \frac{m^2}{\beta^2}(\cos\, \beta \, \Phi )_{\mu}  
\label{lagboson}
\end{eqnarray}
where $\beta^2 = \frac{|4\pi-8e_{\psi}|}{|e_{\psi} k|}$,\, $m^2=\frac{c\, m_{\psi}\mu}{\pi}|4\pi-8e_{\psi}|$,\, and $\epsilon=\mbox{sign}(4\pi-8 e_{\psi})$. Imposing $\epsilon=1$,\, $e_{\psi}< \frac{\pi}{2}$ we get a unitary sine-Gordon theory and a decoupled massless free field.

The bosonized version of the constraints (\ref{equiv}) turns out to be
\begin{eqnarray}
\label{constqua}
<\Psi^{'}| \sqrt{{|4\pi-8e_{\psi}|}/{2\pi}}\,\,\, \partial_{\mu} \rho |\Psi>=0,
\end{eqnarray}
where the $|\Psi>$'s are the space of states of the theory.

The low-energy spectrum of
QCD$_{2}$ has ben studied by means of abelian
\cite{baluni} and non-abelian bosonizations \cite{gonzales, frishman}. In this limit the baryons
of QCD$_{2}$ are sine-Gordon solitons \cite{frishman}. In the large $N$ limit approach (weak $e$ and small $m_{q}$) the SG theory also emerges \cite{salcedo}. 

The low-energy limit of QCD$_{2}$ ($N_{f}=1$) with quarks in the {\sl fundamental} representation of $SU(N)$ is described by the SG theory with (see Appendix)\cite{frishman}
\begin{eqnarray}
\beta = \sqrt{\frac{4\pi}{N}},\,\,\,\,\,\,(N > 1).
\end{eqnarray}

Now, let us introduce in (\ref{lagboson})
a new mass parameter $m'$ by renormal-ordering \cite{abdalla}
\begin{eqnarray}
(\cos\, \beta \Phi)_{\mu} = (\frac{m'}{\mu})^{\frac{\beta^2}{4\pi}} (\cos\, \beta \Phi)_{m'}, 
\end{eqnarray}
then one has
\begin{eqnarray}
{\cal L}_{bos} = - \frac{1}{2 e_{\psi}}  (\partial_{\mu} \rho )^2+ \frac{1}{2} (\partial_{\mu }\Phi )^{2}
+  2\, (m')^2 \,(\cos \beta \, \Phi )_{m'}  
\label{lagboson1}
\end{eqnarray}
where 
\begin{eqnarray}
\label{sgpara}
(m')^2= \Big[ |k e_{\psi}|\, c\, \frac{m_{\psi}}{2 \pi}\,
(\mu)^{\frac{N-1}{N}}\Big]^{\frac{2N}{2N-1}};\,\,\,\,\,\, |ke_{\psi}|=\frac{N}{|\mbox{sign}(e_{\psi})\frac{4N}{|\kappa|} \pm 1|}.
\end{eqnarray}

From (\ref{sgpara}) and the  QCD$_{2}$ parameter $m'$ (\ref{qcdpara}) one can make the identifications $\mu = \frac{e}{\sqrt{2\pi}}; \,\,\,\, m_{\psi}\,\sim\, m_{q}$. An exact relationship between $m_{\psi}$ and $m_{q}$ will be found below. In the large $N$ limit \cite{gonzales, salcedo}:\, $(m')^2 \sim N\, e\, m_{q}$.

On the other hand, QCD$_{2}$ ($N_{f}=1$) with quarks in the {\sl adjoint} representation of $SU(2)$ is described by the SG theory with $\beta^2 = 4\pi$ (see Appendix). This allows us to make the identifications from (\ref{lagboson}) and (\ref{eff2}):\, $m_{\psi} \sim m_{q},\, \mu \sim \Sigma (\sim e)$. 

Let us study the question of confinement of the ``color'' degrees of freedom associated to the field  $\psi$ (see Appendix) in the ATM model by computing the string tension. In a semi-classical analysis \cite{abdalla, armoni}, we put a pair of classical external probe `color' charges $q$ and $-q$ at $L$ and $-L$ described by the static potential $Q_{c}= \alpha [\Theta(x+L)-\Theta(x-L)]$\, ($\alpha$ is a yet unknown factor), in the `color' space direction $T_{ext}^3=$diag$(\lambda_{1},\lambda_{2},..., \lambda_{l},0,0,...,0)$, $\{\lambda_{i}\}$ being the `isospin' components of the representation $R$ under a $SU(2)$ subgroup. Then comparing the vacuum expectation value (v.e.v.) of the Hamiltonian associated to (\ref{lagboson1}) in the presence of an external $(q_{ext})(-q_{ext})$ source with the relevant one in the absence of such a source, we define the string tension in the limit $ L \rightarrow \infty$ \cite{abdalla}
\begin{eqnarray}
\label{tension1}
\sigma =<H>-<H_{0}>,
\end{eqnarray}  
where $H_{0}( H)$ is the Hamiltonian in the absence (presence) of the probe charges. 

Let us examine the `mass' term in (\ref{atm}) with the $SU(N)$ `color' sector of external fields $(\psi^{a})_{\mbox{ext}}$ in the fundamental representation coupled to the Toda field $\varphi$. In ATM type theories the fermions confine in a self-generated potential \cite{chang}, thus coupling the $(\psi_{ext})$'s to $\varphi$ implies some kind of self-coupling in view of the equivalence (\ref{equiv}). From (\ref{nonabel}) we write
$ (\psi^{\dagger\,a}_{L})_{ext}(\psi_{R\, b})_{ext}  = \frac{c \mu}{2\pi} \, \left( e^{iQ_{c}\, T_{ext}^{3}}\right)_{ab} $,         
then the mass term (\ref{massterm}) becomes 
\begin{eqnarray}
\label{masstermext}
k m_{\psi} e_{\psi} {\bar{\psi}}_{a} \,
e^{2i\varphi\,\gamma_5}\, \psi^{a} = \frac{m^2}{2 \beta^2} Tr\left( e^{iQ_{c}\, T_{ext}^{3}}\, e^{2i\varphi}+  e^{-iQ_{c}\, T_{ext}^{3}}\, e^{-2i\varphi}\right).
\end{eqnarray}

Defining the analogue of $\Phi$ in (\ref{linear}) as: $\Phi^{i}_{Q_{c}}= \frac{2}{\beta}(\lambda_{i} Q_{c} + \varphi)$, and from (\ref{linear}) replacing $\varphi$ in terms of the fields $\Phi$ and $\rho$, the mass term in (\ref{lagboson}) can be written as  
\begin{eqnarray}
\label{intq}
\frac{m^2}{\beta^2} \sum_{i=1}^{l}\cos \beta \Phi^{i}_{Q_{c}}\,=\, \frac{m^2}{\beta^2} \,\sum_{i=1}^{l}\Big[\cos \left(  \lambda_{i} Q_{c}-\frac{8 \mbox{sign}(e_{\psi})}{\beta |k|} \Phi+ \frac{8}{\beta |ke_{\psi}|}\sqrt{\pi/2} \,\rho\right) \Big]_{\mu}. 
\label{bosontot}
\end{eqnarray}

The fields $\Phi$ and $\rho$ in (\ref{linear}) inherit from $\varphi$ the symmetries $\rho \rightarrow \rho+\frac{\sqrt{2\pi}}{\beta}\pi n\,\,$  and $\Phi \rightarrow \Phi+\frac{2\pi}{\beta}n$\,\,\, $(n\in \hbox{\sf Z})$, then the theory (\ref{lagboson}) has a degenerate vacua $|0_{n}>$. In order to compute $\sigma$ we need the v.e.v. of the fields, thus we concentrate our attention to one of these vacua, say $|0_{o}>_{\Phi}\, \otimes\, |0_{o}>_{\rho}$. Therefore, in accordance with the constraint ($\ref{constqua}$) we shall set $\rho=0$.

In (\ref{intq}) setting $\rho= 0$ (ATM $\rightarrow$ SG reduction) and $\alpha  =0$ (absence of external charges) we must recover the interaction term $\frac{m^2}{\beta^2} \cos \beta \Phi$ (without the sum in $i$), so we require: $\beta\,=\, \pm [- \frac{8 \mbox{sign}(e_{\psi})}{\beta |k|}]$,\,
where the $\pm$ signs encode the $\Phi \rightarrow \pm \Phi$ symmetry of the SG theory. Then one gets 
\begin{eqnarray}
|\kappa| \,=  \left\{\begin{array}{ll}
 \pm \mbox{sign}(e_{\psi}) 4N, &\,\;\;\mbox{dyn. quarks in the {\sl fundamental} rep. of }\,\,\, SU(N) \label{kappa1}\\
\pm \mbox{sign}(e_{\psi}) 4, &\,\;\; \mbox{dyn. quarks in the {\sl adjoint} rep. of}\,\, SU(2)
\end{array} \right.  
\end{eqnarray}

From (\ref{sgpara}) and (\ref{kappa1}) for the {\sl fundamental} representation one has:
 $|k e_{\psi}|= N/2$,\,\, $|\kappa| = 4N$,\,\,$|e_{\psi}|=\frac{\pi}{4}\approx 0.78$,\,\, therefore\,\, $\frac{m_{\psi}}{4\pi}= m_{q}$. The limit $\frac{|k e_{\psi}|}{4N} \rightarrow \frac{1}{8}$\,\,as $\kappa, \, N\, \rightarrow\,\infty $, is the semi-classical limit of the SG theory ($\beta \rightarrow 0$). In the {\sl adjoint} case one has to compare the coefficients of the `$\cos$' term in (\ref{lagboson}) with its QCD$_{2}$ analogue (\ref{eff2}), thus for $|\kappa|=4$ one has: $\frac{e^{\gamma} m_{\psi} \mu |e_{\psi}|}{\pi^2}= m_{q} \Sigma$ $\rightarrow$ $m_{\psi}\sim m_{q},\, \mu \sim \Sigma (\sim e)$.

In order to describe the chirally rotated mass term in QCD$_{2}$ (\ref{qcdboso}) we set $\alpha = 4\pi \frac{k_{ext}}{k_{dyn}}$ (the case $SU(2)$ requires $2\pi$ instead of $4\pi$) \cite{armoni}. Actually, this is the first order term in the ($\frac{e^2}{M_{q}^2}$) expansion when the external probe charge is viewed as a dynamical field with very large mass $M_{q}$ (see more details in \cite{armoni}). Then the energy v.e.v. in the limit $L \rightarrow \infty$ is: $<H>=- \frac{m^2}{\beta^2} \sum_{i} <\cos (4\pi \lambda_{i} \frac{k_{ext}}{k_{dyn}} + \beta \Phi)>$. Then ($\ref{tension1}$) becomes
\begin{eqnarray}
\label{tension2}
\sigma = \frac{m^2}{\beta^2} \sum_{i=1}^{l}\Big[\left(1- \cos 4\pi \lambda_{i} \frac{k_{ext}}{k_{dyn}} \right) < \cos \beta \Phi > + \sin 4\pi\lambda_{i} \frac{k_{ext}}{k_{dyn}} < \sin \beta \Phi >\Big].  
\end{eqnarray}
Thus the values of $< \cos \beta \Phi >$ and $< \sin \beta \Phi >$ in the SG theory are needed. The {\sl exact} v. e. v. of type $< e^{i a \Phi} >$ (Re$(a)<\frac{\sqrt{2\pi}}{\beta}$) in the SG theory has recently been proposed \cite{zamolodchikov}. The authors studied: ${\cal L}_{SG} = \frac{1}{2} (\partial \Phi)^2 -2 \mu_{o} \cos\beta \Phi$, assuming the normalization
\begin{eqnarray}
\label{normal}
< \cos\beta \Phi(x)\cos\beta \Phi(0)>_{\mu_{o}=0} = \frac{1}{2 |x|^{\frac{\beta^2}{2\pi}}}. 
\end{eqnarray}
From \cite{zamolodchikov} we quote the expectation value for $a=\beta=\sqrt{\frac{4\pi}{N}}$\,\,($N>1$) \cite{note2}
\begin{eqnarray}
< \exp (i \beta \Phi) > &=& C(N)\,\, \mu_{o}^{\frac{1}{2N-1}},\\
\nonumber
C(N)&\equiv&\frac{\frac{2N}{2N-1}}{16 \sin (\frac{\pi}{2N-1})} \left(\frac{\pi \Gamma(1-\frac{1}{2N})}{\Gamma({\frac{1}{2N}})}\right)^{\frac{2N}{2N-1}} \frac{1}{(\Gamma(\frac{N}{2N-1}))^2} \left( \frac{\Gamma(\frac{4N-3}{4N-2})}{4\sqrt{\pi}}\right)^{\frac{1-2N}{N}}\mbox{x}
\\ 
&&\left( \frac{4}{\sqrt{\pi}} \sin(\frac{\pi}{4N-2}) \Gamma(\frac{1}{4N-2})\right)^{\frac{1}{N}}.
\nonumber
\end{eqnarray}
To use this exact result we have to relate $\mu_{o}$ and $m'$. This is done comparing (\ref{normal}) with
\begin{eqnarray}
\nonumber
< \left(\cos\beta \Phi(x)\right)_{m'}\left(\cos\beta \Phi(0)\right)_{m'}> = \cosh\Big[ \beta^2 D(m', |x|)\Big]\,
 \sim\, \cosh \Big[ \frac{\beta^2}{2\pi} (-\gamma -\ln \frac{m'|x|}{2})\Big],
\end{eqnarray}
for small $m'|x|$. We have $(m')^2= c^{\frac{2}{2N-1}}(\mu_{o}^{\frac{2N}{2N-1}})$.

Then the string tension (\ref{tension2}) becomes
\begin{eqnarray}
\label{tension3}
\sigma_{R} =\, \left\{\begin{array}{ll} \frac{2 (m')^2}{(c^{\frac{2}{2N-1}})}\, C(N)\,\,\sum_{i=1}^{l}\left(1- \cos 4\pi \lambda_{i} \frac{k_{ext}}{k_{dyn}}\right),\,\,\, R=\mbox{\sl fundamental rep. of } SU(N)\\
 (m_{q} \Sigma) \,\, \sum_{i=1}^{l}\left(1- \cos 2\pi \lambda_{i} \frac{k_{ext}}{k_{dyn}}\right),\,\,\,\,\,\,\,\,\,\,\,\,\,\,\,\,\,\,\,\,\,R=\mbox{\sl adjoint rep. of } SU(2) \end{array}\right.,
\end{eqnarray}
where $\Sigma$ in $R=adj.$ is the fermion condensate (see Appendix). We propose (\ref{tension3}) as the exact QCD$_{2}$ string tension in the limit $\frac{e}{m_{q}}\rightarrow \infty$. Some comments are in order here:

1) The string tension (\ref{tension}) reproduces qualitatively (\ref{tension3}). Eq. (\ref{tension}) has been derived using a semi-classical average for the bosonized fields in Eq. (\ref{qcdboso}) ($<g>=1$) \cite{armoni}.

2) In the large $N$ limit for $R=\mbox{fund.}$ (\ref{tension3}) takes the form:\\
$\sigma = 2 N c m_{q} \frac{e}{\sqrt{2\pi}}\, \sum_{i} \left(1- \cos 4\pi \lambda_{i} \frac{k_{ext}}{k_{dyn}}\right)\,\,$, which has the same $m_{q}$ and $e$ dependence as (\ref{tension}), except for a  $2\pi N$ factor \cite{note3}. Note that when the dynamical matter is in the fundamental ($k_{dyn}=1$) the string tension vanishes for any external matter. In the $R=${\sl adjoint} of $SU(2)$ case ($k_{dyn}=2$) and external charges in the fundamental $k_{ext}=1$, (\ref{tension3}) reproduces the result of \cite{gross} up to a factor 2. Consider $\vec{\lambda}_{fund}=(\frac{1}{2},-\frac{1}{2},0,0,...,0)$, and  $\vec{\lambda}_{adj}=(1,0,-1)$. 

The $sl(n)$ ATM models may be relevant in the construction of the low-energy effective theories of multiflavor QCD$_{2}$ with the dynamical fermions in the fundamental and adjoint representations, so providing an extension of the picture described above. Notice that in these models the Noether and topological currents (generalizations of (\ref{equiv})) and the generalized sine-Gordon/massive Thirring models equivalences (see Refs. \cite{jmp, bueno}) take place. A work in this direction is under current investigation and will appear elsewhere.

{\sl Acknowledgements}

I thank A. Armoni for his valuable comments about the manuscript which allowed the clarification of certain points, Prof. L.A. Ferreira for collaboration in a previous work which motivated the present investigation, R. Casana for useful conversations, Prof. A. Accioly for encouragement and FAPESP for financial support.

\appendix*

\section{Appendix: The external color charges}
\label{appa}

The equivalence (\ref{equiv}) for multisolitons describes,
 $\varphi = \varphi_{N}$ [$Q_{\mbox{topol}}= N\, \mbox{sign}(e_{\psi})$] and  $\Psi$ $N-$solitons of the SG and MT type, respectively. Asymptotically one can write
\begin{eqnarray}
\label{equiv2}
\frac{1}{2\pi}\epsilon^{\mu\nu} \partial_{\nu} \, \varphi_{N} \approx 
\sum_{a=1}^{N} \frac{1}{\pi} \bar \psi_{a} \gamma^\mu  \psi_{a},
\end{eqnarray}
where the $\psi_{a}$'s are the solutions for the individual localized lowest energy fermion states. In fact, (\ref{equiv2}) encodes the classical SG/MT correspondence \cite{orfanidis}. Thus, the ATM model can accomodate $N-$fermion confined states with internal `color' index $a$ \cite{chang}. 
If we consider $N$ free Dirac fermions $\psi_{a}$ we will have a $SU(N)\mbox{x} U(1)$ symmetry with currents $J^{a}_{\mu}= \bar{\psi}\gamma_{\mu}T^{a}\psi$  and  $J^{\mu}= \bar{\psi}_{a}\gamma^{\mu}\psi^{a}$ where the $T^{a}$'s are the generators of $SU(N)$ in the fundamental representation ($Tr (T^aT^b)=\frac{1}{2} \delta^{ab}$). The $U(1)$ current $J^{\mu}$ was bosonized in (\ref{boson1}).

In order to gain insight into the QCD$_{2}$ origin of the $\psi_{a}$ fields \cite{note4} let us write the mass term in the multifermion sector of ATM theory as 
 \begin{eqnarray}
\label{massterm}
{\bar{\psi}}_{a} \,
e^{2i\varphi\,\gamma_5}\, \psi^{a} = \psi^{\dagger \,a}_{L} \psi_{R\, a}\, e^{2i\varphi} + \psi^{\dagger\, a}_{R} \psi_{L\, a}\, e^{-2i\varphi}.
\end{eqnarray}
The nonabelian bosonization \cite{stone} allows us to write 
\begin{eqnarray}
\label{nonabel}
J^{a}= \frac{-i}{2\pi} Tr(\partial_{-}h h^{\dagger} T^{a}),\,\,\,\bar{J}^{a}= \frac{i}{2\pi} Tr(h^{\dagger}\partial_{+} h T^{a}),\,\,\,
\psi^{\dagger\, a}_{L} \psi_{R\, b}= M \left(h^{a}_{b}\right)_{M} (e^{i\sqrt{4\pi} \phi})_{M},
\end{eqnarray}
where $h$ is a $SU(N)$ matrix field and $x_{\pm}=t\pm x$.
Then (\ref{massterm}) becomes 
\begin{eqnarray}
\label{massboso}
M \left(\mbox{Tr}\, h \, e^{i\beta \Phi} + \mbox{Tr}\, h^{\dagger} \, e^{-i\beta \Phi}\right)_{M},
\end{eqnarray}
where $\Phi= \frac{2}{\beta} (\sqrt{\pi} \phi + \varphi)$ from (\ref{linear}) has been used. The ATM mass term in the multifermion sector, Eq. (\ref{massboso}), must be compared to the corresponding term in the bosonized QCD$_{2}$ in order to identify the fields related to the flavor and color degrees of freedom.  

The bosonized QCD$_{2}$ action ($N_{f}=1$) with a chirally rotated mass term in the {\sl fundamental} and {\sl adjoint} representations can be schematically represented by \cite{armoni}
\begin{eqnarray}
\nonumber
S&=& S_{WZW}[g] + S_{kinetic}[A_{\mu}]- \frac{i k_{dyn}}{4\pi} \int d^2x \mbox{Tr} A_{+}^{a} g\partial_{-} g^{\dagger}\\
\label{qcdboso}
&&+ \frac{1}{2} m_{q} \mu_{R} \int dx^2 \mbox{Tr} \left( g e^{i 4\pi \frac{k_{ext}}{k_{dyn}} T^{3}_{dyn}} +  e^{-i 4\pi \frac{k_{ext}}{k_{dyn}} T^{3}_{dyn}} g^{\dagger}\right), \end{eqnarray}
where $g$ is a $N$x$N$ unitary matrix [ $(N^2-1)$x$(N^2-1)$ orthogonal] for the {\sl fundamental} ({\sl adjoint}) representation, $A_{\mu}$ is the gauge field (the gauge  $A_{-}=0$ was used). When the quarks transform in the {\sl adjoint} rep. the WZW and the interaction terms  must be multiplied by $\frac{1}{2}$ because $g$ is real and represents Majorana fermions ($k_{dyn}=1$ for the fundamental and  $k_{dyn}=N$ for the adjoint reps., respectively), $\mu_{R}$ is to be fixed, and 
\begin{eqnarray}
S_{WZW}[g]=\frac{1}{8\pi}\int d^2x \mbox{Tr}(\partial_{\mu}g\partial^{\mu}g^{-1})+\frac{1}{12\pi}\int d^3y \epsilon^{ijk} \mbox{Tr}(g^{-1}\partial_{i}g)(g^{-1}\partial_{j}g)(g^{-1}\partial_{k}g).
\end{eqnarray}

For quarks in the {\sl fundamental} representation we set  $g = h e^{i\beta \Phi}$ ($\beta =\sqrt{\frac{4\pi}{N}}$), $h\, \epsilon\, SU(N)$, then the mass term becomes
\begin{eqnarray}
\label{massrotated}
 \frac{1}{2} m\, \mu_{fund} \int  d^2x  \mbox{Tr} \left( h e^{i 4\pi \frac{k_{ext}}{k_{dyn}} T^{3}_{dyn}} e^{i \beta \Phi} +  e^{-i 4\pi \frac{k_{ext}}{k_{dyn}} T^{3}_{dyn}} h^{\dagger} e^{-i \beta \Phi} \right), .
\end{eqnarray}

In the strong coupling limit ($\frac{e}{m_{q}} \rightarrow \infty$) the heavy fields can be ignored ($h=1$) after normal ordering at the mass scale $\frac{e}{\sqrt{2\pi}}$. Then (\ref{qcdboso}) becomes the SG model (set $k_{ext}=0$ in (\ref{qcdboso}), i.e. absence of external charges) \cite{frishman}
\begin{eqnarray}
\label{effect}
S_{eff} = \int  d^2x [\frac{1}{2}(\partial_{\mu}\Phi)^2 + 2 (m')^2 \left(\cos \beta \Phi\right)_{m'}],
\end{eqnarray}
where
\begin{eqnarray}
\label{qcdpara}
(m')^2= \Big[N\, c\,\,m_{q}\, (\frac{e}{\sqrt{2\pi}})^{\frac{N-1}{N}}\Big]^{\frac{2N}{2N-1}}
\end{eqnarray}
From (\ref{massboso}) and (\ref{massrotated}) one concludes that the `color' degrees $\psi_{a}$ ($h$ matrix) confined inside the SG solitons correspond to the heavy fields of QCD$_{2}$ which decouple from the light field $\Phi$ at low-energies. 

For quarks in the {\sl adjoint}, one has \cite{smilga}: $g_{ab}= 2 Tr ( T_{a} u T_{b} u^{-1})$, where $u$ is an unitary $N$x$N$ matrix. For $N=2$ and $u=e^{i\sqrt{\pi} \Phi \vec{n}.\vec{\sigma}}$\, ($\vec{n}^2=1$; $\sigma_{a}$, Pauli matrices), the mass term of (\ref{qcdboso}) ($k_{ext}=0$) reproduces exactly: $\cos \sqrt{4\pi} \Phi$, the remaining terms in (\ref{qcdboso}) are the kinetic and derivative interaction terms for the fields $\vec{n}$ and $\Phi$. The kinetic terms do not contribute to the change of the vacuum energy in the presence of the external source \cite{armoni}, and the interaction terms will not contribute in the strong coupling limit. Actually, the change in the vacuum energy is due to the mass term \cite{armoni}. We have the SG model with $\beta^2=4\pi$ \cite{note2} 
\begin{eqnarray}
\label{eff2}
S_{eff}= \int  d^2x [\frac{1}{2}(\partial_{\mu}\Phi)^2 + 2 m_{q} \mu_{adj} \left(\cos \beta \Phi\right)_{\mu_{adj}}].
\end{eqnarray}

When $N=2$, instantons bring about a bilinear fermion condensate (for small $m_{q}$) \cite{smilga}: $2 \mu_{adj} <\cos \sqrt{4\pi} \Phi>= \Sigma\,(\sim e)$.   





\begin{thebibliography}{**}
\bibitem{note2}
According to the SG/MT correspondence the case $\beta=\sqrt{4\pi}\,\,$\,\, is the free fermion point in the SG theory, i.e. the free massive fermions of the MT model.
\bibitem{note4}
In the case of dynamical fermions in the adjoint representation of $SU(2)$ it is possible to construct $SU(2)$ {\sl fund.} rep. from the {\sl adj.} rep. of $SU(2)$ via bosonization \cite{gross}.
\end{thebibliography}
\end{document}